\documentclass[aps,pre,reprint,groupedaddress,amsmath,amsfonts,amssymb,letterpaper]{revtex4-2}
\usepackage{graphicx}
\usepackage{subfigure}
\usepackage{tikz}
\usepackage{mathtools}
\usepackage{appendix}

\usepackage[subtle]{savetrees}% subtle, moderate, extreme
\usepackage{amsmath,amsfonts,amssymb}
\usepackage{graphicx}
\usepackage{subfigure}
\usepackage{mathtools}
\usepackage{makecell}
\usepackage{mathrsfs}
\usepackage{enumitem}

\usetikzlibrary{matrix,decorations.text,decorations.pathmorphing,decorations.markings,arrows,calc,shapes.geometric,patterns,shadows,intersections}
\begin{document}
	\title{Net motion induced by non-antiperiodic\\ vibratory or electrophoretic excitations with zero time average}
	
	\author{Aref Hashemi}
	\email[Email: ]{aref@cims.nyu.edu}
	\affiliation{Courant Institute, New York University, New York, NY, United States}
	\author{Mehrdad Tahernia}
	\affiliation{Department of Information Engineering, The Chinese University of Hong Kong, HK}
	\author{Timothy C. Hui}
        \author{William D. Ristenpart}
        \email[Email: ]{wdristenpart@ucdavis.edu}
        \author{Gregory H. Miller}
	\email[Email: ]{grgmiller@ucdavis.edu}
        \affiliation{Department of Chemical Engineering, University of California Davis, Davis, CA, United States}
    	
%	\date{\today}
	
	\begin{abstract}
	It is well established that application of an oscillatory excitation with zero-time average but temporal asymmetry can yield net drift. To date this temporal symmetry breaking and net drift has been explored primarily in the context of point particles, nonlinear optics, and quantum systems.  Here, we present two new experimental systems where the impact of temporally asymmetric force excitations can be readily observed with mechanical motion of macroscopic objects: (1) solid centimeter-scale objects placed on a uniform flat surface made to vibrate laterally, and (2) charged colloidal particles in water placed between parallel electrodes with an applied oscillatory electric potential. In both cases, net motion is observed both experimentally and numerically with non-antiperiodic, two-mode sinusoids where the frequency modes are the ratio of odd and even numbers (e.g., $2$ $\mathrm{Hz}$ and $3$ $\mathrm{Hz}$). The observed direction of motion is always the same for the same applied waveform, and is readily reversed by changing the sign of the applied waveform, for example by swapping which electrode is powered and grounded. We extend these results to other nonlinear mechanical systems, and we discuss the implications for facile control of object motion using tunable periodic driving forces.
	\end{abstract}
	
	\maketitle
	
	The well known ``ratchet'' effect requires a periodic forcing in a nonlinear system with some sort of broken symmetry \cite{Reimann2002,Hanggi2009}. For example, a sinusoidal driving force in a medium with spatial asymmetry in the resistance to motion yields net drift in the direction of less resistance. Ratchets may also be induced by a variety of different `temporal asymmetries,' where symmetry is broken in the periodic excitation rather than the physical medium \cite{Flach2000,Denisov2002,Ustinov2004,Denisov2014}. As discussed in detail by Denisov et al. \cite{Denisov2014}, periodic excitations that are not `shift-symmetric' (also known as `antiperiodic') can induce net drift. Prior work has considered and experimentally corroborated this type of `temporally induced' ratchet in the context of point particles \cite{Flach2000,Yevtushenko2001,Denisov2002,Dukhin-Dukhin2005}, as well as optical \cite{Schiavoni2003,Jones2004,Gommers2005,Gommers2006,Struck2012,Eckardt2017} and quantum \cite{Denisov2007a,Denisov2007b} lattice systems. Although the theory predicts that temporally asymmetric force excitations will also cause net motion of macroscopic objects, to date experimental evidence for this claim is scarce. The goal of this article is to introduce two new experimental systems where temporal ratchets are easily induced with macroscopic and easily visualized objects.
 
	To provide context, we begin by presenting a streamlined derivation of the requirements for a temporal asymmetry in the context of mechanical motion (consistent with prior results, e.g., \cite{Denisov2014}). Consider an object subject to a generic periodic force excitation $f(t)$ with period $2\tau$ and a resistance to motion of the form $G(v)$, where $v$ is the object velocity. By Newton's second law, the equation of motion is
	\begin{equation}
	m\frac{dv}{dt}=f(t)-G(v).
	\label{Eq:Newton}
	\end{equation}
	We restrict focus to excitations with zero time-average,
	\begin{equation}
	\langle f \rangle = \frac{1}{2\tau}\int_0^{2\tau} f(t)dt = 0.
	\label{Eq:timeavg}
	\end{equation}
	Furthermore, we are interested in situations where the resistance to motion does not favor one direction over another, so necessarily $G(v)$ is an odd function of $v$. This restriction excludes the wide variety of ratchet-like problems where net motion is induced simply because motion occurs more easily in one direction versus another \cite{Reimann2002}.

        \begin{figure*}[t]
		\centering
		\setlength{\belowcaptionskip}{-10pt}
		\resizebox{1\linewidth}{!}{\includegraphics{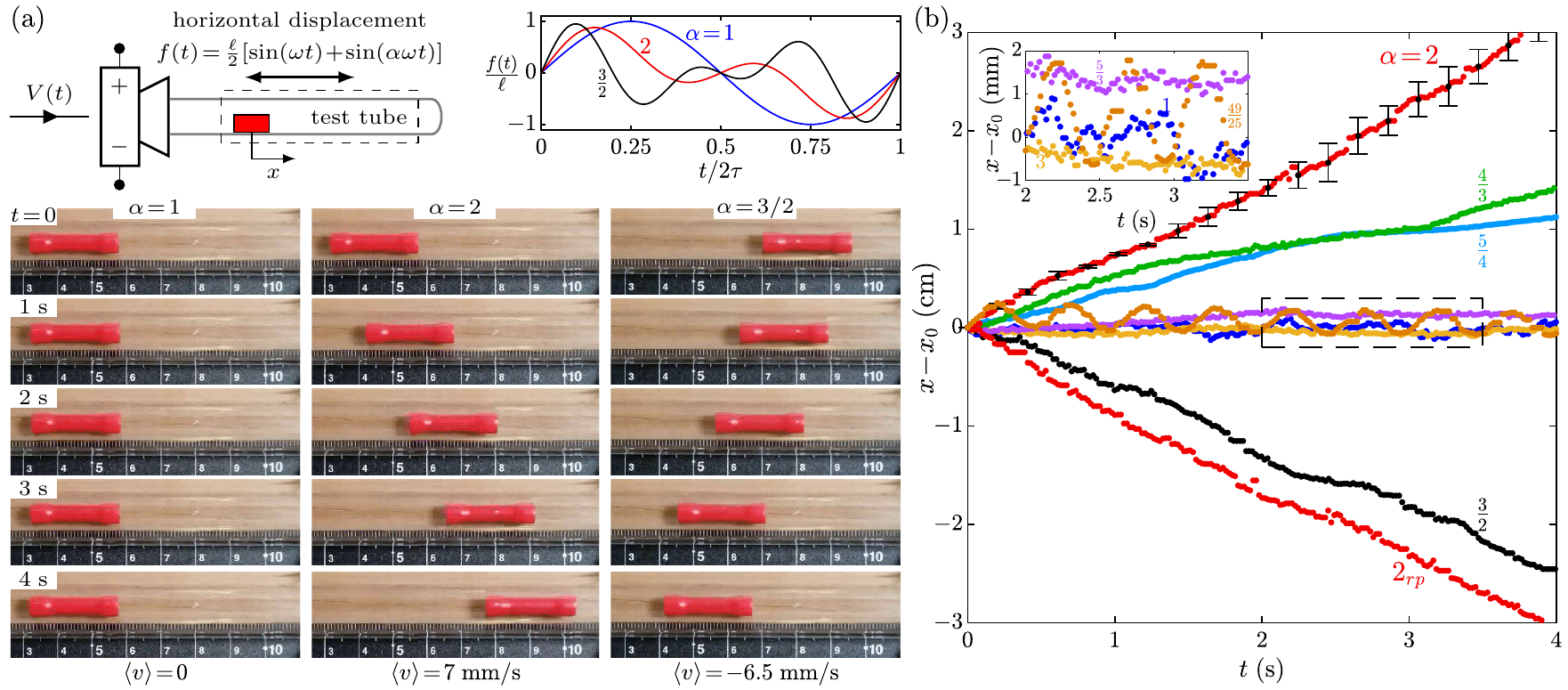}}
		\caption{Experimental evidence for deterministic ratchets in the solid-solid friction problem. (a), top: Schematic diagram of the system; an object is placed in a test tube that is glued to the diaphragm of a speaker. A two-mode sound wave is fed to the speaker with base frequency $\omega/2\pi=50$ $\mathrm{Hz}$ (see Appendix for further details). (a), bottom: Time lapse photos of the object dynamics for $\alpha=1$, $2$, and $\tfrac{3}{2}$, corresponding to applied frequency modes of 50 Hz, 50 and 100 Hz, and 50 and 75 Hz, respectively. (b): Object location versus time for different $\alpha$ values. Representative error bars are two standard deviations of the mean of at least three trial replicates. Also see Supplementary Video 1.}
		\label{Fig:Fig_1}
	\end{figure*}
        
	In a spatially-symmetric odd-$G(v)$ system, let $v$ be the unique solution to Eq.~(\ref{Eq:Newton}) with initial condition $v(0)=0$. In this circumstance, $-v(t)$ is the solution to the reverse polarization excitation $-f$: the functional $\psi$, where $v(t)=\psi(f,t)$, is odd in $f$.
	
	To make further progress, we must next specify something about $f(t)$. In particular, note that some periodic functions are also antiperiodic, which is the term used to describe any periodic function with period $2\tau$ that, for all $t$, obeys the relationship
	\begin{equation}
	f(t+\tau)=-f(t).
	\label{Eq:antiperiodic}
	\end{equation}
	That is, the second half of an antiperiodic waveform is equal to the negative of the first half \cite{Freire2013}. All single-mode sinusoids are antiperiodic, but multimodal sinusoids can be antiperiodic or non-antiperiodic depending on the frequency modes. A subtle but important feature of non-antiperiodic functions is that $f(t)$ and the reverse polarity function $-f(t)$ are intrinsically different in the sense that no choice of time lag maps one onto the other.
	
	If a driving force satisfies the antiperiodic condition Eq.~(\ref{Eq:antiperiodic}), there is an important consequence for the force balance Eq.~(\ref{Eq:Newton}). Specifically, because the functional $\psi(f,t)$ is odd, if $f(t)$ is antiperiodic, we have
	\begin{equation}
	v(t+\tau)=\psi(f,t+\tau)=\psi(-f,t)=-v(t),
	\end{equation}
	so $v(t)$ is also antiperiodic. In the long time limit, when the
	influence of initial conditions is negligible, a time lag does not alter the time-average solution:
	%, i.e., $\langle v(t+\tau)\rangle=\langle v(t)\rangle$, we obtain
	\begin{equation}
	\langle v(t)\rangle=-\langle v(t)\rangle,
	\end{equation}
	which can only be satisfied if $\langle v(t)\rangle =0$. Therefore, antiperiodic driving forces yield zero net motion if the resistance to motion $G(v)$ is odd. More detailed arguments giving rise to the same conclusion are presented in ref. \cite{Denisov2014}. 
	
	As a demonstrative example, a two-mode excitation of the form $f(t)=\tfrac{1}{2}[\sin(t)+\sin(\alpha t)]$, which has period $2\tau=2\pi/\mathrm{gcd}(1,\alpha)$, is antiperiodic if
	\begin{equation}
	\sin(t+\tau)+ \sin\left(\alpha(t+\tau)\right) = -\sin(t)-\sin(\alpha t),
	\end{equation}
	which requires $\tau = (2j+1)\pi$ and $\alpha \tau = (2k+1)\pi$ for
	integer $j$ and $k$, and
	\begin{equation}
	\alpha = \frac{2k+1}{2j+1}.
	\end{equation}
	Two-mode excitations are antiperiodic provided that the mode ratio $\alpha$ can be expressed as the ratio of two odd numbers.
	
	What happens if the force excitation has zero time-average but is non-antiperiodic? The preceding proof showed that antiperiodic force excitations necessarily yield zero time-average velocities, but it leaves open the possibility that non-antiperiodic excitations might yield non-zero net velocities. Clearly, whether a non-zero time-average occurs in that situation will also depend sensitively on the nature of $G(v)$; for example, if $G(v)$ is a linear function of $v$, it is straightforward to show that $\langle v\rangle$ must be zero even for non-antiperiodic functions, provided $\langle f\rangle=0$. If $G(v)$ is a nonlinear odd function, however, there is no \emph{a priori} reason why $\langle v\rangle$ must be zero for non-antiperiodic excitations. We now present experimental evidence demonstrating that non-antiperiodic force excitations indeed can yield net motion of macroscopic objects.
	
	\begin{figure*}[t]
		\centering
		\setlength{\belowcaptionskip}{-10pt}
		\resizebox{1\linewidth}{!}{\includegraphics{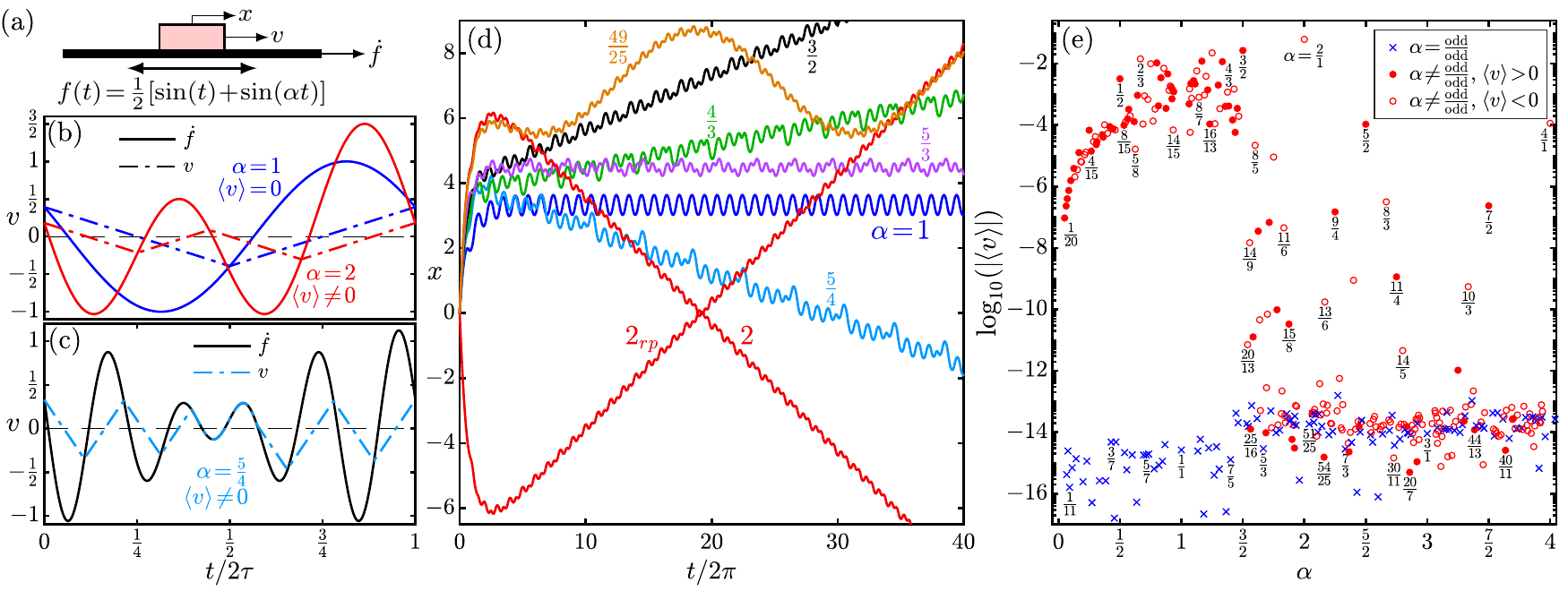}}
		\caption{Dynamic response of an object placed atop a solid surface (solid-solid friction) to a two-mode displacement excitation of the surface $f(t)=\tfrac{1}{2}[\sin(t)+\sin(\alpha t)]$ (dimensionless form). (a): schematic diagram of the problem. (b, c): representative solutions to the harmonic object and surface velocities ($v$ and $\dot{f}$, respectively) versus time, for different $\alpha$ values. (d): object location, $x$, versus time for different $\alpha$ values. The case $\alpha_{rp}$ denotes the response due to the reverse polarization excitation $f(t)=-\tfrac{1}{2}[\sin(t)+\sin(\alpha t)]$. (e): absolute value of the harmonic time-average object velocity, $\mid\!\langle v\rangle\!\mid$, versus $\alpha$. Here, time is scaled by the inverse angular frequency $1/\omega$, and the dimensionless period is $2\tau=2\pi/\mathrm{gcd}(1,\alpha)$, where $\mathrm{gcd}(1,\alpha)$ denotes the greatest common divisor of $1$ and $\alpha$. The solution is considered `harmonic' if it is invariant between different periodic intervals. Dimensionless parameters: $\lambda_s=0.5,\;\lambda_k=0.25$.}
		\label{Fig:Fig_2}
	\end{figure*}

	We first consider a solid-solid friction system where an object is placed on a vibrating surface (Fig.~\ref{Fig:Fig_1}). An object (here a red plastic cylinder) is placed in a glass test tube attached perpendicularly to a standard dynamic speaker of the kind typically found in television sets. A two-mode digital sound wave is fed to the speaker, generating a periodic back and forth movement of the speaker diaphragm, which, in turn, induces a two-mode lateral vibration of the test tube with displacement $f(t)=\tfrac{\ell}{2}[\sin(\omega t)+\sin(\alpha \omega t)]$, where $\ell$ is the oscillation amplitude, and the ratio of frequency modes $\alpha$ is a rational number. Importantly, the tube providing the frictional driving force remains stationary on average, i.e., the excitation $f$ clearly has zero time-average, $\langle f\rangle=0$. Despite the zero time-average vibration, however, the experimental observations indicate that the system behavior depends sensitively on $\alpha$. The time lapse photos in Fig.~\ref{Fig:Fig_1} and Supplementary Video 1 show that for a unimodal frequency of 50 Hz ($\alpha=1$), the plastic cylinder remains stationary. Likewise, for a two-mode waveform with 50 Hz and 150 Hz ($\alpha=3$), the cylinder also remains stationary. In contrast, application of a waveform with 50 Hz and 100 Hz ($\alpha=2$) immediately causes the cylinder to displace rightward at $\langle v\rangle=7$ mm/s. A waveform with 50 Hz and 75 Hz ($\alpha=3/2$) caused the cylinder to displace leftward at $\langle v\rangle=-6.5$ mm/s. Tests for several different values of $\alpha$ (Fig.~\ref{Fig:Fig_1}(b)) showed that net drift of varied magnitude was observed when $\alpha$ was even or a ratio that included an even number (e.g., $\alpha=\tfrac{3}{2}$, $\tfrac{4}{3}$, $\tfrac{5}{4}$, or $2$), but that only oscillatory motion with no net displacement was observed if $\alpha$ was an odd number or a ratio of two odd numbers (e.g., $\alpha=1$, $\tfrac{5}{3}$, $\tfrac{49}{25}$, $3$). Notably, in each case reversing the polarity of the applied waveform (accomplished by switching the leads to the speaker) induced motion in the opposite direction but with equivalent speed. One such reverse polarity trial is shown in (Fig.~\ref{Fig:Fig_1}(b)), denoted as $\alpha=2_{rp}$; note that the average velocity is equal and opposite to that induced by the original polarity.
	
	To rule out the possibility that there was something unusual about the plastic cylinder, we repeated the experiment with a variety of different solid objects, including pebbles, metal washers, and coffee beans. In each case we observe no net drift for two-mode waveforms with frequency modes that can be expressed as the ratio of odd numbers, and controllable motion to the left or right for waveforms with frequency modes that include an even number, the direction of which was always swapped upon reversing the polarity. To mitigate any possible influence of confined sound waves (the human eye cannot follow the vibratory motion but the speaker emits an audible hum, cf. Supplementary Video 1), we replaced the enclosed glass test tube with a flat metal plate. Again, similar behavior was observed on the flat plate, suggesting it is indeed the frictional interaction between the object and the vibrating surface that induces motion.
	
	We emphasize that, unlike the classical Feynman--Smoluchowski ratchet \cite{Reimann2002}, the solid-solid friction system shown in Fig.~\ref{Fig:Fig_1} has no spatial asymmetry. Instead, the phenomenon appears to stem solely from a time-symmetry break in the excitation. Furthermore, the observed net motion is \emph{deterministic}; the direction of motion remains the same for different trials. There is a rich literature on nonlinear oscillations and frictional interactions with vibrating systems; see for example the text by Nayfeh and Mook \cite{Nayfeh1979}. Although this prior work does not describe net motion of the sort presented in Fig.~\ref{Fig:Fig_1}, it does provide a theoretical framework to develop a model for the behavior. We assume that the object and substrate interaction can be described in terms of the two standard coefficients of friction used with Coulomb's law of friction: the coefficient of static friction $\mu_s$ when the object and substrate do not move relative to each other, and the coefficient of kinetic friction $\mu_k$ when the object slides along the substrate at a different velocity. Typically these coefficients are not equal, and they depend sensitively on the composition of the two surfaces. We neglect drag force in the air and assume the only lateral force acting on the object is the frictional force of the substrate moving below it with position described by the imposed periodic displacement waveform $f(t)$. In this case, the dimensionless equations of motion for the object are
\begin{equation}
\dot{v}=
\begin{cases}
  \begin{alignedat}{2}
&\ddot{f}\quad&&\text{if}\quad v=\dot{f}\;\text{and}\;\mid\ddot{f}\mid<\lambda_s,\\
&-\lambda_k\mathrm{sgn}(v-\dot{f})\quad&&\text{otherwise}.
  \end{alignedat}
\end{cases}
\label{Eq:SS_Model}
\end{equation}
Here, $\mathrm{sgn}$ denotes the sign function; the length and time dimensions are scaled by $\ell$ (vibration amplitude) and $1/\omega$ (inverse base angular frequency); we define $\lambda_s=\mu_sg/\ell\omega^2$ and $\lambda_k=\mu_kg/\ell\omega^2$ as the dimensionless static and kinetic friction coefficients, respectively; and the dimensionless two-mode vibration is $f(t)=\tfrac{1}{2}[\sin(t)+\sin(\alpha t)]$. Essentially, whenever the substrate accelerates sufficiently slowly, the object is simply carried along at the same velocity, but if the substrate accelerates too quickly, the object cannot keep up and slides along at a smaller velocity. According to Eq.~(\ref{Eq:SS_Model}), the system behavior is governed by the three dimensionless parameters $\alpha$, $\lambda_s$, and $\lambda_k$. The behavior is independent of mass because both the inertial term and the frictional force are linearly proportional to the object mass (see Appendix for additional details).

	\begin{figure*}[t]
		\centering
		\setlength{\belowcaptionskip}{-10pt}
		\resizebox{1\linewidth}{!}{\includegraphics{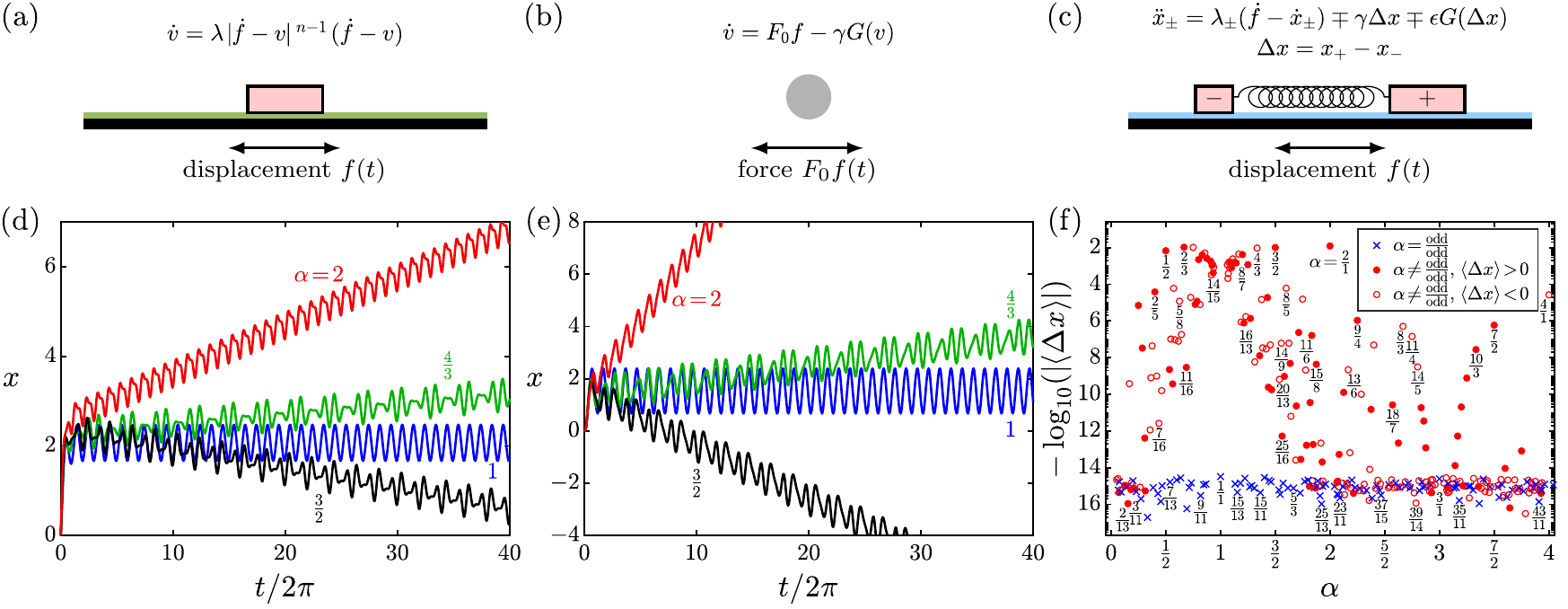}}
		\caption{Existence of deterministic ratchet-like behavior in various nonlinear dynamical systems. Each vertical figure panel corresponds to one problem. (a, b, c): schematic diagrams of various nonlinear dynamical systems under a two-mode excitation $f(t)=\tfrac{1}{2}[\sin(t)+\sin(\alpha t)]$, with their dimensionless governing equations. (a): an object subject to the drag of a non-Newtonian liquid film on an excited surface. (b): an object subject to a force excitation and a nonlinear drag $G(v)=v^3$. (c): a pair of asymmetric objects connected by a nonlinear spring ($G(\Delta x)=\Delta x[(\Delta x-1)^{-2}+(\Delta x+1)^{-2}]$), and subject to the drag of a Newtonian liquid film on an excited surface. (d, e): time evolution (numerically evaluated) of the object location ($x$ versus $t$) for different $\alpha$. (f): time-average dimensionless distance between the two masses for different $\alpha$. Dimensionless parameters: (a, d): $\lambda=0.5,\;n=1.5$; (b, e): $F_0=1,\;\gamma=1$; (c, f): $\lambda_+=10,\;\lambda_-=0.1,\;\gamma=0.1,\;\epsilon=0.1$.}
		\label{Fig:Fig_3}
	\end{figure*}
	
	Note that the model is nonlinear by virtue of the $\mathrm{sgn}$ function since there is a discontinuity in the direction of the imposed force (so it is not Lipschitz-continuous), but it is readily solved by numerical methods (see Appendix for details). Here we focus on the impact of $\alpha$. During a single time period for $\alpha=1$ (a unimodal vibration), we see that even after the direction of the substrate switches from positive to negative (solid blue curve, Fig.~\ref{Fig:Fig_2}(b), $\alpha=1$, near $t/2\tau=0.1$), the object continues to drift in the positive direction as it decelerates until finally switching directions (dashed blue curve, $\alpha=1$, near $t/2\tau=0.25$). The object keeps accelerating in the negative direction, until the magnitude of the substrate velocity falls below the magnitude of the object velocity (intersection of solid and dashed blue lines near $t/2\tau=0.5$), at which point the substrate begins accelerating the object in the opposite direction. Similar behavior is observed for the two-mode waveform with $\alpha=2$, albeit with different periods of acceleration and deceleration (red curves, $\alpha=2$, Fig.~\ref{Fig:Fig_2}(b)). For a more complicated waveform with $\alpha=\tfrac{5}{4}$ (Fig.~\ref{Fig:Fig_2}(c)), there are periods where the frictional force does not exceed the static friction, so the object just moves in tandem with the substrate (approximately near $0.4<t/2\tau<0.6$).
	
	Inspection of the corresponding object positions versus time shows that the model indeed yields net drift for certain waveforms (Fig.~\ref{Fig:Fig_2}(d)). After a brief transient period, the time-average velocity for $\alpha=1$ is zero (dark blue curve, $\alpha=1$). Similarly, for the waveform $\alpha=\tfrac{5}{3}$, after the transient stage, the object ends up in a different position, but its average velocity is again zero (purple curve, $\alpha=\tfrac{5}{3}$). In contrast, for $\alpha=2$, the object displaces quickly in the negative direction, while applying the reverse polarity of the same waveform causes the object to displace quickly in the positive direction (red curves, $\alpha=2$). Qualitatively, this behavior is strikingly similar to the experimental observations in Fig.~\ref{Fig:Fig_1}. Systematic computational investigation of a wide variety of different values of $\alpha$ indicates that qualitative differences in the direction and magnitude of the motion occur with seemingly small differences in $\alpha$ (Fig.~\ref{Fig:Fig_2}(e)). There is no clear pattern to the distribution of positive and negative velocities, but one trend is clear: whenever the frequency modes are the ratio of odd numbers, the average velocity is zero (i.e., in the limit of numerical noise). In contrast, values of $\alpha$ that include an even number typically (but not always) induce a nonzero velocity. Furthermore, we stress that the results in Fig.~\ref{Fig:Fig_2}(e) are representative. The direction of motion for a constant $\alpha$ (i.e., retaining the same spatial structure of the waveform) can be reversed by changing the frequency and amplitude of the excitation. This is a reminiscent of the observed current reversals in rocking ratchets \cite{Cubero2010,Wickenbrock2011}.

        A key implication of the general argument is that non-antiperiodic driving forces can induce net motion for any system governed by the generic Eq.~(\ref{Eq:Newton}), not just the solid-solid frictional system explored in Figs.~\ref{Fig:Fig_1} and \ref{Fig:Fig_2}. To test this idea, we developed numerical models for several different nonlinear dynamical systems.  First, we replace the solid-solid friction with the drag of a non-Newtonian fluid film (Fig.~\ref{Fig:Fig_3}(a)). Here, the source of nonlinearity is the nonlinear shear stress from the fluid, rather than the solid-solid friction. The system is spatially symmetric, nonlinear, and its solution (the object velocity, $v(t)$) changes sign upon changing the polarity of the excitation, and hence, has all the requirements needed for temporally induced ratchets. Indeed, our representative numerical results (Fig.~\ref{Fig:Fig_3}(d)) show that, similar to the solid-solid frictional problem, the object stays stationary on average for antiperiodic vibrations ($\alpha=1$), but drifts for non-antiperiodic ones ($\alpha=\tfrac{4}{3}$, $\tfrac{3}{2}$, and $2$). Another example is an isolated sphere subjected to a two-mode periodic force excitation and a nonlinear drag $G(v)$ (Fig.~\ref{Fig:Fig_3}(b)). (A familiar practical example is a colloid translating through a non-Newtonian fluid in response to a two-mode force.) Note that, again, we are interested in nonlinear drag terms that do not favor a direction over another (i.e., $G(v)$ is odd in $v$). We used a variety of odd nonlinear drags such as $G(v)=v^3,\;\;\mathrm{sgn}(v),\;\;\mid\!v\!\mid\!v,\;\;\sinh{(v)}$ and observed the same qualitative behavior; as demonstrated in Fig.~\ref{Fig:Fig_3}(e), the system behaves like a ratchet under non-antiperiodic forces, and induces a net drift of the object (see $\alpha=\tfrac{4}{3}$, $\tfrac{3}{2}$, and $2$).
		
	Our modeling indicates that temporally-induced ratchets also occur with spring forces, provided the spring force is non-Hookean (Fig.~\ref{Fig:Fig_3}(c)). Here a pair of asymmetric masses are connected by a nonlinear spring, and placed atop an excited surface coated with a Newtonian fluid film (linear shear). Similar to the system in Fig.~\ref{Fig:Fig_3}(b), we tested various odd nonlinear spring forces. Shown here is a spring force model that imposes two `solid walls' at displacements $\Delta x=1$ and $\Delta x=-1$, which ensure that the spring does not elongate to more than double of its resting value nor compress to negative values. Figure~\ref{Fig:Fig_3}(f) shows the scaled time-average distance between the two masses for different $\alpha$ values. We note that for antiperiodic vibrations (blue crosses in Fig.~\ref{Fig:Fig_3}(f)) the time-average of $\Delta x=x_+-x_-$ is zero. This result indicates that the average distance between the two masses remains equal to the resting value. However, $\langle \Delta x\rangle$ can be nonzero otherwise (red circles in Fig.~\ref{Fig:Fig_3}(f)). Depending on $\alpha$, the masses stay farther apart ($\langle\Delta x\rangle>0$) or closer ($\langle\Delta x\rangle<0$) than their resting condition. We emphasize that here, unlike the previous examples, the masses do not drift. It is instead the time-average separation between the two masses (not their velocity) that exhibits a ratchet-like behavior.

        It is worth mentioning that our solid-solid friction system (Figs.~\ref{Fig:Fig_1} and \ref{Fig:Fig_2}) and the systems described in Fig.~\ref{Fig:Fig_3}(a) and (b) are all overdamped. In a dynamical system, oscillations occur due to conversion of energy between different forms (e.g., between kinetic and potential energy). It is a well know fact that a system with a one-dimensional phase space (one type of energy) cannot oscillate \cite{Strogatz1994}. In mathematical terms, oscillations are impossible in systems with a governing equations which are reducible to a first order differential equation. The mass spring problem illustrated in Fig.~\ref{Fig:Fig_3}(c) is, however, more complicated. The spring can store energy and hence, the system can be either overdamped or underdamped, depending on the system parameters.
	
\begin{figure*}[t]
		\centering
		\setlength{\belowcaptionskip}{-10pt}
		\resizebox{1\linewidth}{!}{\includegraphics{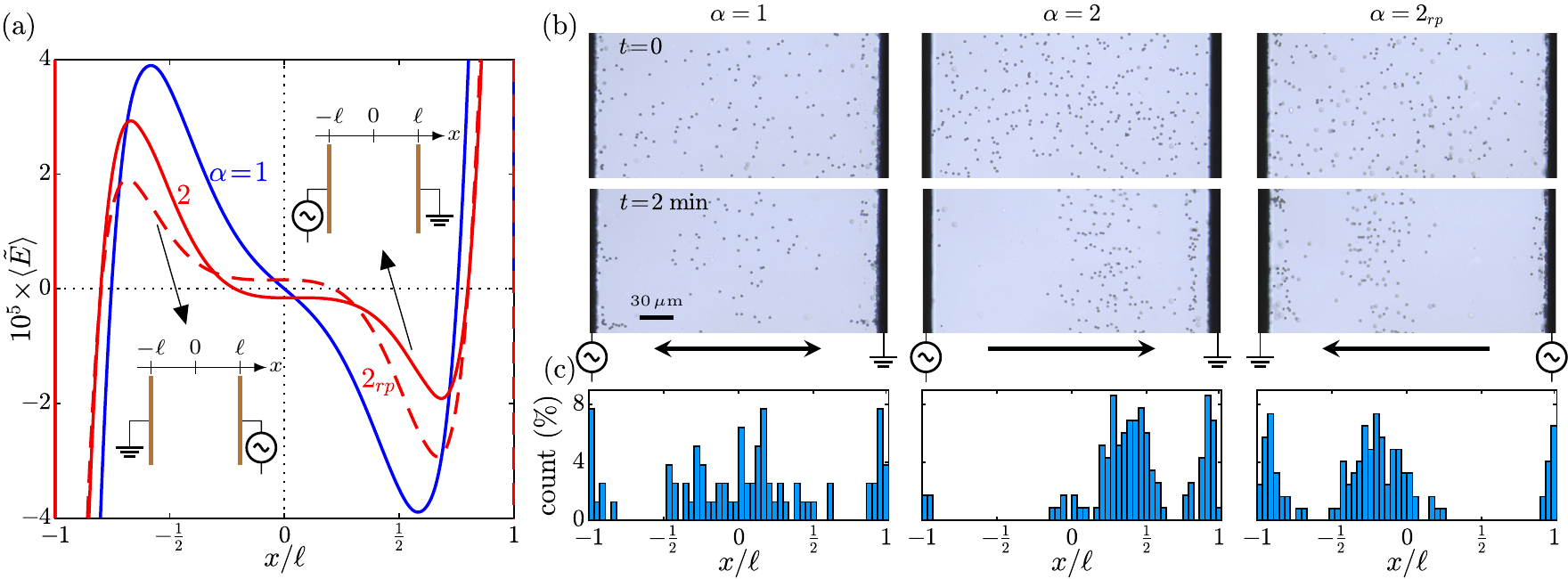}}
		\caption{Deterministic ratchets in the induced steady electric field between two parallel electrodes, placed at $\pm\ell$, and under a two-mode potential excitation $\phi(t)=\tfrac{1}{2}\phi_0(\sin(\omega t)+\sin(\alpha\omega t))$. (a) (Numerical solution): spatial distribution of the dimensionless time-average electric field, $\langle \tilde{E}\rangle=2\ell \langle E\rangle/\phi_0$, at the micron scale, for $\alpha=1$, $2$, and $2_{rp}$. The case $\alpha_{rp}$ denotes the response due to the reverse polarization excitation (the powered and grounded electrodes are swapped). Parameters: $\phi_0=10\phi_T$, $\omega/2\pi=50$ $\mathrm{Hz}$, $2\ell=30$ $\mu\mathrm{m}$, $1$ $\mathrm{mM}$ NaOH solution. (b, c) (Experimental evidence): electrophoresis of charged colloidal particles due to asymmetric rectified electric field (AREF). (b): cluster of the colloids at $t=0$ (no field) and $t=2$ $\mathrm{min}$ (equilibrium conditions) for $\alpha=1$, $2$, and $2_{rp}$. The black arrows show the drift direction of the colloids. (c): the corresponding histograms of the percent particle count after $t=2$ $\mathrm{min}$. Parameters: $\phi_0=4$ $\mathrm{V}$, $\omega/2\pi=2$ $\mathrm{Hz}$, $2\ell=270$ $\mu\mathrm{m}$, $0.01$ $\mathrm{mM}$ NaOH solution, $2\text{--}\mu\mathrm{m}$ sulfonated polystyrene particles. Also see Supplementary Videos 2--4 for $\alpha=1$, $2$, and $2_{rp}$, respectively.}
		\label{Fig:Fig_4}
	\end{figure*}

	As a final test, we investigated the electrophoretic motion of colloidal particles in response to time-varying electric potentials. Micron-scale colloids in water are well known to move back and forth electrophoretically in response to application of an AC electric field. The resulting velocity of the colloids, however, is strongly coupled with the response of the dissociated electrolytes present in the water to the time varying field \cite{Russel1991}. The motion of the colloids cannot be reduced to Eq.~(\ref{Eq:Newton}). Nonetheless, our numerical and experimental results indicate that application of a non-antiperiodic electric potential does give rise to a temporally-induced ratchet (Fig.~\ref{Fig:Fig_4}).

	Here we consider the classical electrokinetic problem of a 1-1 binary electrolyte confined between two planar, parallel, electrodes at $x\!=\!\pm\ell$. A two-mode potential $\phi(t)=\tfrac{1}{2}\phi_0(\sin(\omega t)+\sin(\alpha\omega t))$ (or $-\phi(t)$ for the reverse polarization) is applied on the electrode at $x\!=\!-\ell$, and the electrode at $x\!=\!\ell$ is grounded. In continuum theory, the system behavior is governed by the Poisson-Nernst-Planck model, a notoriously nonlinear and coupled system of equations \cite{Russel1991}. The nonlinearity of the problem stems from the electromigration of ions due to the time varying, multimodal, electric field present within the liquid. Recent work focused on the single-mode potentials ($\alpha=1$) established that oscillatory electric potentials induce a nonzero time-average electric field within the electrolyte, $\langle E\rangle$, referred to as asymmetric rectified electric field (AREF) \cite{Aref2018,Aref2019,Aref2020SM}, provided the ions present have unequal mobilities. The single-mode AREF is antisymmetric in space, and is identically zero at the midplane ($\alpha=1$, solid blue curve in Fig.~\ref{Fig:Fig_4}(a)), meaning the steady field is independent of which electrode is powered or grounded. In contrast, when $\alpha=2$, a non-antisymmetric $\langle E\rangle$ is induced, with a non-zero electric field at the midplane. Here, swapping the powered and grounded electrodes \emph{does} alter the system; notably, the sign/direction of $\langle E\rangle$ at the midplane changes ($\alpha=2$ and $2_{rp}$ in Fig.~\ref{Fig:Fig_4}(a)).

	Our experimental observations accord with the numerical results. We evaluated the action of the induced $\langle E\rangle$ on a cluster of randomly dispersed colloids in the electrolyte (Figs.~\ref{Fig:Fig_4}(b) and (c)). As demonstrated in Fig.~\ref{Fig:Fig_4}(b) and Supplementary Video 2, the colloids are equally attracted to either electrode for $\alpha=1$. When $\alpha=2$, however, we observe a significant asymmetry in the movement of the colloids (see also Supplementary Video 3). For the representative conditions studied here, Fig.~\ref{Fig:Fig_4}(b), and the corresponding histogram of the equilibrium distribution in Fig.~\ref{Fig:Fig_4}(c), clearly show that the colloids moved preferentially towards the right electrode (the grounded one). Swapping the powered and the grounded electrodes reversed the drift direction (cf. $\alpha=2_{rp}$ in Figs.~\ref{Fig:Fig_4}(b) and (c), and Supplementary Video 4). (Please see Appendix for details of the electrokinetic experiments.)
	
	To summarize, the preceding theoretical, experimental, and numerical results all strongly indicate that non-antiperiodic, zero-time-average, driving forces can induce net motion in isotropic media. Several questions, however, remain unanswered. Perhaps the most obvious question is: \emph{which direction will the object move?} Clearly, net motion is induced, but at present we have not identified analytical or heuristic criteria to relate the nature of the imposed non-antiperiodic waveform to nonlinearities in the equation of motion. Vidybida and Serikov \cite{Vidybida1985} present a theory for small and Lipschitz-continuous restorative forces $G(v)$ (e.g., Fig.~\ref{Fig:Fig_3}(b)), and for $\alpha=2$ specifically, which is not applicable for our frictional system. Ultimately, the direction of motion must be controlled by a subtle interplay between the waveform and the nonlinear terms in the equation of motion. For the electrokinetic system, the coupled nonlinear equations with multiple length and time scales tremendously complicate the interpretation. More importantly, the model does not explicitly reduce to the generalized force-resistance problem given by Eq.~(\ref{Eq:Newton}). But, even for the vibratory frictional system with a much simpler equation of motion, it is unclear why the object drifts to the left versus the right. Furthermore, it is unclear as to what determines the magnitude of the response for different non-antiperiodic waveforms. Our results for a variety of problems suggest that $\alpha=2$ induces the strongest temporal ratchet (e.g., highest drift velocity). Meanwhile, some non-antiperiodic waveforms tend to induce a near-zero response (cf. Fig.~\ref{Fig:Fig_2}(e)). Why some waveforms yield strong motion, while others do not, remains unclear. Additionally, we have not considered the existence of so called `hidden' symmetries in our systems. In particular, overdamped dynamical systems exhibit symmetries that are unidentifiable by standard symmetry analyses \cite{Cubero2016}.

        Although we focused here on two-mode sinusoids, the theory is not limited to the two-mode excitations; our results suggest any zero-time-average and non-antiperiodic excitation (e.g., sawtooth waves, pulse waves, triangle waves) can yield temporally-induced ratchets. Likewise, other types of periodic driving forces (magnetic, hydrodynamic, acoustic) might give rise to net motion if they are non-antiperiodic. The results presented here serve as a framework to consider temporally-induced ratchets in these more complicated systems.
        
	\noindent\textbf{Supplementary Material.} The experimental observations are available via the supplementary videos 1--4.
	
	\noindent\textbf{Acknowledgments.} This material is based upon work partially supported by the National Science Foundation under Grants No. DMS-1664679 and No. CBET-2125806.
	
	\noindent\textbf{Conflict of Interest.} The Authors declare that there is no conflict of interest.
	
	\noindent\textbf{Data Availability.} The numerical and experimental data are available from the corresponding author upon reasonable request.

\renewcommand{\thesection}{A}
\renewcommand{\theequation}{A\arabic{equation}}

\section*{Appendix}

\subsection*{Numerical solutions}
The following loop solves the solid-solid friction problem (Fig.~\ref{Fig:Fig_2}). Initially, $v=\dot{f}$, and then,
\begin{enumerate}[label=\roman*),noitemsep]
\item $v=\dot{f}$ as long as $\lvert\ddot{f}\rvert<\lambda_s$.
\item Once $\lvert\ddot{f}\rvert>\lambda_s$, sliding starts: $\dot{v}=-\lambda_k\mathrm{sgn}(v-\dot{f})$, until $v=\dot{f}$ again.
\item Go to step i.
\end{enumerate}
The time $t$ is updated in each step as well. Whenever $t$ increases by $2\tau$ is considered a cycle of the solution. We repeat the cycles until a harmonic solution is achieved. Let $v_k(t)$ with $t\in[2(k-1)\tau,2k\tau]$ denote the solution in the $k^{\text{th}}$ cycle. The solution is considered harmonic if $\lVert v_k-v_{k-1}\rVert<\epsilon$, where $\epsilon$ is a tolerance. We also check $\lvert \langle v_k\rangle-\langle v_{k-1}\rangle\rvert<\epsilon$.
		
The toy problems in Fig.~\ref{Fig:Fig_3} are solved by the Runge--Kutta 4\textsuperscript{th} order method. The same criteria is used for the harmonic solution check. Details of the numerical solution to the nonlinear electrokinetic problem (Fig.~\ref{Fig:Fig_4}(a)), and the corresponding consistency checks are provided elsewhere \cite{Aref2018,Aref2019}.
		
\subsection*{Solid-solid friction model}
Here we derive the solid-solid friction model given by Eq.~(\ref{Eq:SS_Model}) of the main manuscript. Consider an object of mass $m$, and velocity $v$, on top of a flat plate of mass $M$ and velocity $V$ (cf. Fig.~\ref{Fig:Fig_2}(a)). A horizontal force $F$ is applied on the plate.
		
The coupling frictional force $F_c$ (on the object) is due to static and kinetic friction between the object and the plate ($\mu_s$ and $\mu_k$ are the corresponding friction coefficients). Equations of motion for the two masses are
\vspace{-0.25cm}
\begin{align}
m\dot{v}&=F_c,\\
M\dot{V}&=F-F_c.
\end{align}
		
The coupling force can be determined for when $v=V$ and $\dot{v}=\dot{V}$ (i.e., the object and the plate are moving in tandem) as $F_c=mF/(m+M)$. Note that this velocity matching condition is maintained if $v=V$ initially and $\mid F_c\mid<\mu_s F_N$. Here $F_N=mg$ is the normal force. Otherwise, the object starts sliding (or keeps sliding) with $F_c=-\mu_kF_N\mathrm{sgn}(v-V)$. Hence, one can write the Newton's second law for the object as
		
\begin{widetext}
\begin{equation}
	m\dot{v}=
	\begin{cases}
	\begin{alignedat}{3}
	&\frac{mF}{m+M}\quad&&\text{if}\quad v=V\;\text{and}\;\left\lvert\frac{mF}{m+M}\right\rvert<\mu_sF_N,\quad&&\text{(velocity matched)}\\
	&-\mu_kF_N\mathrm{sgn}(v-V)\quad&&\text{otherwise},\quad&&\text{(sliding)}
	\end{alignedat}
	\end{cases}
	\label{Eq:SS_Model_Dimensional_General}
	\end{equation}
			
Now let $M\gg m$ and $F=M\ddot{f}$, with $f$ as the lateral displacement of the surface, to obtain
\begin{equation}
	\dot{v}=
	\begin{cases}
	\begin{alignedat}{3}
	&\ddot{f}\quad&&\text{if}\quad v=\dot{f}\;\text{and}\;\mid \ddot{f}\mid<\mu_sg,\quad&&\text{(velocity matched)}\\
	&-\mu_kg\mathrm{sgn}(v-\dot{f})\quad&&\text{otherwise},\quad&&\text{(sliding)}
	\end{alignedat}
	\end{cases}
	\label{Eq:SS_Model_Dimensional}
\end{equation}
which is the dimensional form of Eq.~(\ref{Eq:SS_Model}).
\end{widetext}

\subsection*{Solid-solid friction experiments}
An object (a wire splice connector) is placed in a glass test tube of length 15 cm and outer diameter 18 mm (IWAKI TE-32 PYREX), that is glued to the diaphragm of a used television speaker ($R=8$ $\mathrm{\Omega}$). A two-mode sound wave, created by MATLAB, enters a generic class D amplifier. The amplified current is then fed to the speaker as an excitation. The sound actuator behaves linearly, i.e., its movement is linearly proportional to the passing current. Harmonic oscillations of the diaphragm translate to a one-dimensional displacement excitation of the tube. Note that the tube geometry restricts the object to move in one dimension. A digital camera is used to record the object dynamics at 60 frames per second. Note that the passing current, and consequently, the displacement amplitude, are kept sufficiently low to ensure a linear behavior of the sound actuator. As a result, the movement of the tube itself is not easily discernible.
		
\subsection*{Electrokinetic experiments}
The experimental setup consists of a microchannel constructed using two flat sheets of polydimethylsiloxane (PDMS), that were separated by two $16$ $\mu\mathrm{m}$ thick 304 stainless steel plates spaced $270$ $\mu\mathrm{m}$ apart. (In Fig.~\ref{Fig:Fig_4}(b), the electrodes (stainless steel plates) are $270$ $\mu\mathrm{m}$ apart, the depth of the cell is $16$ $\mu\mathrm{m}$ (through the page), and the point of view is through the PDMS sheet.) The channel had a total length of $15$ $\mathrm{mm}$. Two polyethylene tubes of $0.58$ $\mathrm{mm}$ inner diameter were inserted into the top PDMS layer to introduce and remove the fluid. Copper tape was used to connect the stainless-steel sheets to the powered and grounded wires. The device was then sealed using epoxy around the edges and fixed in place over a glass substrate using clamps.
		
A $0.01$ $\mathrm{mM}$ NaOH solution (conductivity, $\sigma=2$ $\mu\mathrm{S}/\mathrm{cm}$) was prepared using DI water ($18.2$ $\mathrm{M\Omega.cm}$), and $2\text{--}\mu\mathrm{m}$ diameter fluorescent sulfonated polystyrene particles were added at a volume fraction of $1\times10^{-4}$ to the solution. The colloidal suspension was washed three times by centrifugation and resuspension, and then injected into the microchannel using a syringe pump (PHD 2000, Harvard apparatus). Once the flow inside the microchannel was stable and the particle density appeared uniform, a function generator (Agilent 33220A) was used to apply a sum modulated field of $4$ $\mathrm{V_{pp}}$ (Volts peak-to-peak) at $2$ $\mathrm{Hz}$ and $4$ $\mathrm{V_{pp}}$ at $4$ $\mathrm{Hz}$. A digital camera mounted on an optical microscope (Leica DM2500 M) was used to record the particle behavior at $15$ frames per second. After two minutes, the field was removed and the channel was then flushed for a minute. The powered electrode was changed by physically exchanging the wire leads on the device, upon which the same field was then applied. For the unimodal case ($\alpha=1$), an $8$ $\mathrm{V_{pp}}$ at $2$ $\mathrm{Hz}$ field was applied using the same procedure. Furthermore, when $\alpha=1$, swapping the powered and grounded electrodes had no significant impact on the system behavior.

%% \bibliographystyle{apsrev4-1}%apsrev4-1%apsrmp4-1%plain
%% \bibliography{../manuscript-revised}


\end{document}